\documentstyle[12pt]{article}
\begin{document}
\thispagestyle{empty}
\noindent
DO-TH 93/29   \\ 
OHSTPY-HEP-T-93-011   \\
\hfill September 1993 \\
\vspace{1cm}
\begin{center}
  \begin{Large}
  \begin{bf}
A PARTON MODEL FOR INCLUSIVE SEMILEPTONIC B MESON DECAYS

     \end{bf}
 \end{Large}
\end{center}
  \vspace{1cm}
   \begin{center}
C.\ H.\ Jin$^a$\footnote{Supported by Bundesministerium f\"ur Forschung und
Technologie, 05-6DO93P, Bonn, FRG and by the CEC Science Project
$n^{o}$ SC1-CT91-0729.},
W.\ F.\ Palmer$^b$\footnote{Supported in part by the US
            Department of Energy under contract DOE/ER/01545-605.}
            and E.\ A.\ Paschos$^a$$^1$\\
      \vspace{0.3cm}
        $^a$Institut f\"ur Physik\\  Universit\"at Dortmund\\
        D--44221 Dortmund, Germany\\
  \vspace{0.3cm}
        $^b$Department of Physics, The Ohio State University \\
        Columbus, Ohio 43210, USA\\
\end{center}

\vfill
\begin{abstract}
The parton model for semileptonic B meson decays is studied with special
attention to the decay distributions. We find that the spectra show
dramatic variations when we introduce cuts on the hadronic energy or
invariant mass of hadrons. Results for both $b\rightarrow u$ and
$b\rightarrow c$ decays are presented. The detailed spectra may help to
separate the two types of decays.
\end{abstract}
\newpage
The decays of B-mesons have been studied extensively and have been
very useful in extracting properties of the weak interactions. In
particular, the $b\rightarrow u$ transition is still an active field
of investigation. On the experimental side there is a large effort
hindered by the difficulty of identifying the decays of B-mesons to
light quarks. This difficulty of identifying $b\rightarrow u$ events
forced investigators to concentrate on the end-point spectrum of the
electron energy or study exclusive decays\cite{exp:tobe}.
In this letter we present
an extended study of the semileptonic B-decays. We are
interested in the decay
\begin{equation}
B\rightarrow X_{u}+e^{-}\bar \nu
\end{equation}
for which we wish to give explicit distributions in several kinematic
variables. With the decay spectrum completely determined by a single
parameter in the distribution function, it may be
compared to the competing process in $B\rightarrow X_c +e^-\bar \nu$.
With the help of these distributions
it should be possible to test characteristics of the decay
products with various kinematical cuts, and, in particular, collect events
typical
for this process.  The hope is that with the results presented here the
experimental analyses could incorporate many more events than those
confined to
the endpoint region of the electron spectrum.

\bigskip
There are three models
for the inclusive decays of reaction (1). The early model of Altarelli
et al.\cite{alt:np}, to be denoted as ACM, uses a spectator model
and a distribution of quarks within the B-meson described by the Fermi
motion of the spectator quark. This approach treats the phase space
effects correctly, but is rather crude, as the authors
state\cite{alt:np}, because it depends on an unknown distribution
function for the spectator quark. The model pictures the overall decay
as the disintegration of the B-meson into the spectator quark plus the
decay products of the heavy quark.

\bigskip
The second approach\cite{bar:np,bar:zp} uses the parton model in an
infinite
momentum frame (IMF).  The probability of finding a b-quark in a B-meson
carrying a fraction $x$ of the mesons momentum in the IMF is given by
the distribution function $f(x)$. The distribution function $f(x)$
has the functional form suggested by theoretical
arguments\cite{suzuki:pl,bj:pr,balay:sov}
and it peaks at large values of $x$.
The kinematics for the decay of a heavy to
a light quark involve the correlation of two currents at short
distances, where the incoherence of the decay products is justified.
This model has recently been improved and applied to
double differential inclusive decay distributions in reference
\cite{jin:np}
where a detailed treatment of the kinematics and the allowed physical
regions
can be found. A special feature of the parton model is the distribution of
b quark in the B-meson, which provides a continuum spectrum for the mass of
the recoiling hadrons.
Here we only summarize the main results and point out
new features before turning to applications.

\bigskip
A third model, suggested very recently\cite{bigi:prl},
improves the free quark model by including $1/m_{Q}$ corrections from the
heavy quark theory. These corrections turn out to be rather small.
Obviously, the three models incorporate
 dynamical corrections
in a different manner. In the parton model dynamic corrections are included
through the experimentally determined distribution function and
contributions from two scaling variables $x_{+}$ and $x_{-}$, to be
described below.

\bigskip
In the parton model\cite{bar:np,bar:zp} the decay kinematics are different
from those in deep inelastic  scattering. As a consequence,
the energy-momentum conserving
$\delta$-function has now two roots, that is, it gives two scaling
variables
\begin{equation}
x_{\pm}=\frac{q_{0}\pm \left| \vec q \right|}{M_{B}}=\frac{q_{\pm}}{M_{B}}
\end{equation}
which are the light-cone variables of the current-momentum. Thus we
have again scaling of the distribution function but now in terms of
two variables.
We define the
kinematic variables:  $P_{B}$=  momentum of B-meson;
$P_{e}, E_e$            =    momentum, energy of electron;
$P_{\nu}$               =    momentum of neutrino;
$P_{X}$                 =    momentum of hadrons;
$q=P_{e}+P_{\nu}$       =    momentum of current; and
$M_X$                   = invariant mass of the
final hadronic system. In terms of these the decay is in general
 defined as:
\begin{equation}
d\Gamma= \frac{G^{2}\left| V_{ub}\right|^{2}}{(2\pi)^{5}M_{B}}L^{\mu\nu}
W_{\mu\nu}\frac{d^{3}P_{e}}{2E_{e}}\frac{d^{3}P_{\nu}}{2E_{\nu}} .
\end{equation}
The leptonic tensor has the simple form
\begin{equation}
L^{\mu\nu}= 2(P_{e}^{\mu}P_{\nu}^{\nu}+P_{\nu}^{\mu}P_{e}^{\nu}-
g^{\mu\nu}P_{e}\cdot P_{\nu}+i\varepsilon^{\mu\nu}
\ _{\alpha\beta}P_{e}^{\alpha}
P_{\nu}^{\beta}) .
\end{equation}
The hadronic tensor is defined in analogy to deep inelastic scattering,
\begin{equation}
W_{\mu\nu}(P_{B},q)= -g_{\mu\nu}W_{1}(q^{2},q\cdot P_{B})+
\frac{P_{B\mu}P_{B\nu}}{M_{B}^{2}}W_{2}(q^{2},q\cdot P_{B})-
i\varepsilon_{\mu\nu\alpha\beta}\frac{P_{B}^{\alpha}q^{\beta}}{M_{B}^{2}}
W_{3}(q^{2},q\cdot P_{B})+\cdots
\end{equation}
where the integration over the hadronic variables has already been performed.
  After carrying out the
integrations over $x$ we arrive at the structure functions\footnote{Note
that the structure functions in the present article are different from those
in \cite{bar:zp}.}
\begin{eqnarray}
W_{1}(q^{2},q\cdot P_{B}) & = & 2\lbrack f(x_{+}) - f(x_{-})\rbrack ,  \\
W_{2}(q^{2},q\cdot P_{B})/M_{B}^{2} & = & \frac{4}{M_{B}\left| \vec q \right|}
\lbrack x_{+}f(x_{+}) + x_{-}f(x_{-})\rbrack  , \\
W_{3}(q^{2},q\cdot P_{B})/M_{B}^{2} & = & -\frac{2}{M_{B}\left| \vec q
 \right|}\lbrack f(x_{+}) + f(x_{-})\rbrack .
\end{eqnarray}
and hence the $b\rightarrow u$ triple differential
decay rate
\begin{equation}
\frac{d\Gamma}{dE_{e}dq^{2}ds}= \frac{G^{2}\left| V_{ub}\right|^{2}}
{8\pi^{3}M_{B}}\frac{q_{0}-E_{e}}{\left| \vec q \right|}\lbrace
x_{+}f(x_{+})(2 E_{e}-M_{B}x_{-}) + (x_{+} \leftrightarrow x_{-})\rbrace .
\end{equation}
This formula is simple and shows the dependence on the two light-cone
variables $x_{\pm}$ .

\bigskip
The distribution function $f(x)$ can be taken from references
\cite{suzuki:pl,bj:pr,balay:sov}.
Its functional form is very similar to the fragmentation
function of a b quark into a B-meson. It has also been argued, on physical
grounds, that the distribution and fragmentation functions for heavy mesons
are the same (reciprocity relation)\cite{bar:np,bro:pr,peterson:pr}.
The fragmentation function was
measured in several experiments and can be
represented\cite{peterson:pr,chrin:zp} as
\begin{equation}
f(x)= N\frac{x(1-x)^{2}}{\lbrack (1-x)^{2}+\varepsilon_{p}x\rbrack^{2}}
\end{equation}
with $\varepsilon_{p}$ a parameter and $N$ the normalization factor. We
will use three values
$\varepsilon_{p} = 0.003, 0.006$.

\bigskip
Finally, we must comment on the kinematic regions where the model is valid.
Two criteria must be fulfilled.

\bigskip
(i) The decay involves the correlation of two currents. We are more
confident in applying the model when this distance between the two currents
is close to the light-cone or at short distances.

\bigskip
(ii) The model should apply in the region where many particles are
produced. Thus, we must avoid the edge of the phase space where one or
two pions
are produced.

\bigskip
To sum up, we feel rather confident in using the parton model for the
decay $B\rightarrow X_{u}+e^{-}\bar \nu$ away from the edges of the phase
space. For the decay $B\rightarrow X_{c}+e^{-}\bar \nu$,
we are less confident,
because the distances involved are larger.
Furthermore,
for the decays $B\rightarrow X_{c}+e^{-}\bar\nu$ we must keep the c-quark
mass, which modifies equation (9),
and it makes a difference whether we choose the minimum value of
the final
hadronic mass as $m_{c}$ or $M_{D}$, as we will show in figures 1b and 2a.

\bigskip
As we mentioned, one of the purposes of this work is to present detailed
spectra which can eventually be compared with experiments. The simplest
parameter to measure is the electron energy. In addition one may be able
to measure the total hadronic energy $E_{X}= M_{B}-q_{0}$. In Fig.~1a
we show the $b \rightarrow u$
double differential decay rate in $E_{e}$ and $q_{0}$. The
spectra show a striking dependence of  $q_{0}$. Most of the events occur for
$0.3 M_{B}\leq E_{X}\leq 0.5 M_{B}$. For smaller hadronic energies the
spectra shift to higher electron energies, $E_{e}\geq 2.0 \ GeV$. This
correlation of events may help to isolate $b\rightarrow u$ events.

\bigskip
For comparison we also calculated the distribution for the
$B\rightarrow X_{c}+e^{-}\bar\nu$ decay. In Fig.~1b we show the double
differential decay rate in $E_{e}$ and $q_{0}$. Comparing with Fig.~1a we
note that this channel runs out of events at  $E_{e}=2.25$ $GeV$ and the events
above 2 $GeV$ are very few. By making a cut in $q_{0}>M_{B}-M_{D}=0.65M_{B}$
there are no events left for $b\rightarrow c$. This can be used as another
criterion for isolating $b\rightarrow u$ events.

\bigskip
The integrated spectrum $d\Gamma/dE_{e}$ is shown in Fig.~2a for two
fragmentation parameters $\varepsilon_{p}$. In the same figure we show the
corresponding spectra for $b\rightarrow c$. For these curves we have chosen
$\left|V_{cb}\right| = \left| V_{ub}\right|$. It is clear that the
$b\rightarrow c$ spectrum is softer and a very small fraction lies at
$E_{e}>2.0$ $GeV$. We show two spectra for the $b\rightarrow c$ decays, by
varying the minimum value of the final hadronic mass. We have chosen two
values $(M_{X})_{min}=m_{c}=1.5 \ GeV$ and $(M_{X})_{min}=M_{D}=1.86 \ GeV$.
We notice that the total decay rate varies considerably when we vary
$(M_{X})_{min}$. We feel that the final quark mass should be replaced by the
running charm quark mass\footnote{This topic is still under investigation.}.
As a result, the choice of $(M_{X})_{min}$ influences the determinations of
$\left|V_{cb}\right|$.

\bigskip
For comparison we calculated the $b \rightarrow u$ spectrum $d\Gamma/dE_{e}$
in the ACM
model\cite{alt:np}. We show in Fig.~2b the ACM and parton spectra
together. (In these spectra we include QCD radiative corrections which
appear as a multiplicative factor, as described in \cite{jin:np} after
the work of \cite{alt:np,corbo:np}.) We note that the parton spectrum is
lower and falls off less steeply than in the ACM model. Consequently, the
values for $V_{ub}$ extracted from the end-point energy will be {\it larger}
in the parton model with the experimental fragmentation function of
eqn (10).  Of course we can reproduce the free quark model by modifying the
fragmentation function so that it has the limit $\delta(1-x)$ when the width
parameter tends to zero, but this is not what is being measured for the
fragmentation function
\cite{peterson:pr,chrin:zp,balay:sov}. This means that the semileptonic
rate of a free $b$ quark is substantially greater than that of a
$b$ quark confined in a $B$ meson. This is analogous to deep inelastic
scattering, where moments of the distribution functions were measured to be
smaller than one\cite{par:pr}.
Thus the spectra in Fig. 2b differ
not only in shape but also in overall scale.

\bigskip
The double differential $b \rightarrow u$ decay rate $d\Gamma/dE_{e}dM_{X}$
depends strongly
on the invariant mass of the hadrons $M_{X}$. We show in Fig.~3
$d\Gamma/dE_{e}dM_{X}$ for various values of $M_{X}$. The decay spectrum is
larger in the range $0.2M_{B}\leq M_{X}\leq 0.4M_{B}$. The characteristic
feature again is that the spectra shift to larger values of $E_{e}$ as
$M_{X}$ decreases. After integration over $E_{e}$ the sum of the curves
gives the mass distribution, which agrees with mass distributions
published before\cite{bar:np} and is similar but not identical with the
curves in ref. \cite{ba:pl}.

\bigskip
The parton model provides an interesting alternative for
analysing inclusive semileptonic decays in important kinematical regions
where
the decay is dominated by short distance physics.  We have produced decay
spectra for a variety of interesting and physically accessible observables
which probe the decay dynamics in a much more complete way than the simple
electron spectrum.  These spectra are obtained from a simple and compact
formula (eq.~(9)) based on a one parameter parton model.  As high statistics
data become available this model may play an increasingly
important role in separating $b \rightarrow u$ from $b \rightarrow c$ decays
and lead to a better determination of the $V_{ub}/V_{cb}$. In fact, the
distribution function $f(x)$ can be measured, in principle, in the
$b\rightarrow u$ decays and then be used as an input to calculate the
$b\rightarrow c$ spectra.

\bigskip
CHJ wishes to thank Deutscher Akademischer Austauschdienst (DAAD) for
financial support.
WFP wishes to thank the North Atlantic Treaty Organization for a Travel
Grant.  EAP wishes to acknowledge the hospitality of
The Ohio State University, where this work was initiated.

\newpage
\centerline{\bf Figure Captions}

\bigskip
\bigskip

\bigskip
1a. The $b\rightarrow u$ double differential decay rate $d\Gamma/dE_{e}dq_{0}$
vs $E_{e}$ for various values of $q_{0}$ in the rest frame of the B meson.
We set $m_{u}= 0, M_{\pi}= 0, M_{B}= 5.3 \ GeV, \alpha_{s}= 0$ and
$\varepsilon_{p}= 0.006$.

\bigskip
1b. The $b\rightarrow c$ double differential decay rate $d\Gamma/dE_{e}dq_{0}$
 vs $E_{e}$ for various values of $q_{0}$ in the rest frame of the B meson.
For the solid lines we use $m_{c}= 1.5 \ GeV, M_{B}= 5.3 \ GeV,
(M_{X})_{min}=M_{D}= 1.86 \ GeV, \alpha_{s}= 0$
and $\varepsilon_{p}= 0.006$. The dashed lines have the same parameters except
for $(M_{X})_{min}=m_{c}=1.5 \ GeV$. For $q_{0}=0.4M_{B}$ the two curves
coincide.

\bigskip
2a. $d\Gamma/dE_{e}$ for semileptonic B meson decays in the rest frame of the
B meson. The values for the parameters are $m_{u}= 0, m_{c}= 1.5 \ GeV,
M_{B}= 5.3 \ GeV, M_{\pi}= 0, M_{D}= 1.86 \ GeV, \alpha_{s}= 0$ and various
values of
$\varepsilon_{p}$ as shown. For the $b\rightarrow c$ decays the low and high
lines correspond to $(M_{X})_{min}=M_{D}$ and $(M_{X})_{min}=m_{c}$,
respectively.

\bigskip
2b. The $b\rightarrow u$ spectrum in the ACM model and the parton model.
Both curves include QCD radiative corrections with $\alpha_{s}= 0.24$.

\bigskip
3. The $b\rightarrow u$ double differential decay rate $d\Gamma/dE_{e}dM_{X}$
vs $E_{e}$ for various values of $M_{X}$ in the rest frame of the B meson.
The values for the parameters are the same as in Fig.~1a.

\end{document}